\documentclass[conference]{IEEEtran}

\usepackage{amsmath}
\usepackage{balance}
\usepackage[noadjust]{cite}
\usepackage{threeparttable}

\ifCLASSINFOpdf
  \usepackage[pdftex]{graphicx}
\else
\fi
\usepackage{array}

\usepackage{multirow}

\begin{document}
%
\title{A Single-Cycle MLP Classifier Using Analog MRAM-based Neurons and Synapses}



\author{\IEEEauthorblockN{Ramtin Zand}
\IEEEauthorblockA{Department of Computer Science and Engineering, University of South Carolina, Columbia, SC 29208. (ramtin@cse.sc.edu)
}
}
\maketitle

\begin{abstract}
In this paper, spin-orbit torque (SOT) magnetoresistive random-access memory (MRAM) devices are leveraged to realize sigmoidal neurons and binarized synapses for a single-cycle analog in-memory computing (IMC) architecture. First, an analog SOT-MRAM based neuron bitcell is proposed which achieves $12\times$ reduction in power-area-product compared to the previous most power- and area-efficient analog sigmoidal neuron design. Next, proposed neuron and synapse bitcells are used within memory subarrays to form an analog IMC-based multilayer perceptron (MLP) architecture for the MNIST pattern recognition application. The architecture-level results exhibit that our analog IMC architecture achieves at least two and four orders of magnitude performance improvement compared to a mixed-signal analog/digital IMC architecture and a digital GPU implementation, respectively, while realizing a comparable classification accuracy.


\end{abstract}



%
\IEEEpeerreviewmaketitle

\section{Introduction}

In-memory computing (IMC) has attracted considerable attention in recent years as a hardware accelerator for artificial neural networks (ANNs) \cite{inmemory-fan,in-memory-wong}. The main objective of the IMC architectures as alternatives for von-Neumann architectures is avoiding the processor-memory bottleneck to realize an energy-efficient and area-sparing computation. To achieve this goal, various techniques have been investigated from using 3D integration technology \cite{in-memory-mutlu} to leveraging beyond-CMOS memristive devices \cite{in-memory-xie}. Recently, various resistive technologies have been proposed to be used within IMC architectures such as resistive random access memory (ReRAM) \cite{in-memory-xie}, phase-change memory (PCM) \cite{in-memory-PCM}, and magnetoresistive random-access memory (MRAM) \cite{inmemory-fan}. 

Most of the previous IMC approaches operate in the digital domain \cite{inmemory-fan, in-memory-pimball}, meaning that they leverage resistive memory crossbars to implement Boolean logic operations such as XNOR/XOR within memory subarrays, which can be utilized to implement multiplication operation in binarized neural networks \cite{xnor-net}. While digital IMC approaches provide important energy and area benefits, they are not fully leveraging the true potential of resistive memory devices that can be realized in the analog domain. On the other hand, mixed-signal analog/digital IMC architectures \cite{in-memory-PCM, Neuromemristive2020} leverage the resistive crossbars to compute multiply and accumulation (MAC) operation in O(1) time complexity, however, they still require transferring data to digital processors to compute activation functions. Thus, in addition to the energy that is consumed to transfer data between processor and memory, signal conversion blocks are required to convert data from analog to digital domain and vice versa, which can lead to considerable energy overheads. In this paper, we use spin-orbit torque (SOT)-MRAM technology to implement both synapses and neurons within analog IMC subarrays that can be concatenated to form a multilayer perceptron (MLP) classifier that operates in a single clock cycle.

\begin{figure}
\centering
\includegraphics[scale=0.4]{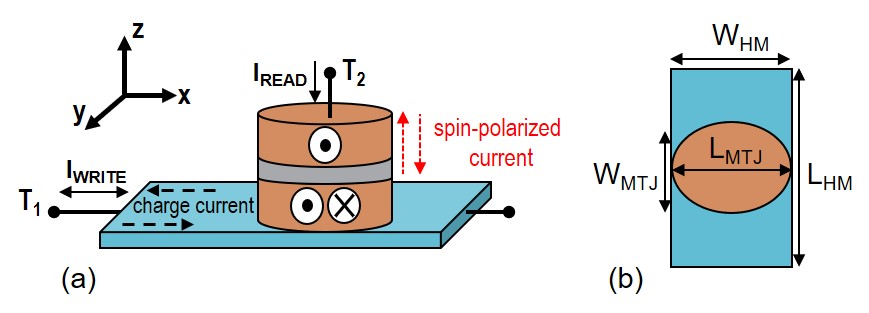}
\caption{(a) SOT-MRAM cell. Positive current along $+x$ induces a spin injection current $+z$ direction. The injected spin current produces the required spin torque for aligning the magnetic direction of the free layer in $+y$ directions, and vice versa. (b) SOT-MRAM Top view.}
\label{fig:sotmram}
\end{figure}

\section{SOT-MRAM based Neurons and Synapses}
Fig. \ref{fig:sotmram} shows a simplified structure of a SOT-MRAM cell including a magnetic tunnel junction (MTJ) with two ferromagnetic (FM) layers, which are separated by a thin oxide layer. MTJ has two different resistance levels, which are determined according to the angle ($\theta$) between the magnetization orientation of the FM layers. The resistance of the MTJ in parallel (P) and antiparallel (AP) magnetization configurations can be obtained using the following equations \cite{Zhang2012CompactModeling}:

\begin{equation} 
\label{EqR} 
\small 
\begin{aligned}
R(\theta) = & \frac{2R_{MTJ}(1 + TMR)}{2 + TMR ( 1 + \cos\theta)}\ \\= 
& \begin{cases} 
R_P=R_{MTJ}, & \theta=0  \\ 
R_{AP}=R_{MTJ}(1+TMR), & \theta=\pi 
\end{cases} 
\end{aligned}
\end{equation}

\begin{equation} \small \label{EqTMR} TMR(T,V_b)= \frac{TMR_0/100}{1+(\frac{V_b}{V_0})^2}\    \end{equation}

\noindent where $R_{MTJ} = \frac{RA}{Area}$, in which the resistance-area product (RA) value of the MTJ depends on the material composition of its layers. TMR is the tunneling magnetoresistance, which relies on temperature (T) and bias voltage ($V_b$). $V_0$ is a fitting parameter, and $TMR_0$ is a material-dependent constant.

In the MTJ structure, the magnetization direction of electrons in one of the FM layers is fixed (pinned layer), while the electrons' directions in the other FM layer (free layer) can be switched. In \cite{Liu2012}, Liu et al. have shown that passing a charge current through a heavy metal (HM) generates a spin-polarized current using the spin Hall Effect (SHE), which can switch the magnetization direction of the free layer, as described in Fig. \ref{fig:sotmram}. The ratio of the generated spin current to the applied charge current is normally greater than one leading to an energy-efficient switching operation \cite{zandTVLSI, zandTNANO}. Herein, we use (\ref{EqR}) and (\ref{EqTMR}) to develop a Verilog-A model of the SOT-MRAM device using the parameters listed in Table \ref{table:sheparameters} \cite{Zhang2012CompactModeling}. The SOT-MRAM model is utilized along with the 14nm HP-FinFET PTM library to implement the neuron and synapse circuits that are described in the following.


\begin{table}
\centering
\footnotesize
\caption{Parameters of the SHE-MRAM device \cite{Zhang2012CompactModeling}.}
\label{table:sheparameters}
\begin{tabular}{c c c} \hline  
{\bf Parameter} & {\bf Description} & {\bf Value} \\ \hline
\multirow{2}{*}{$MTJ_{Area}$} &  \multirow{2}{*}{$l_{MTJ}\times w_{MTJ} \times \frac{\pi}{4}$} & \multirow{2}{*}{$50nm \times 30nm \times \frac{\pi}{4}$}  \\ 
		{}&{}&{} \\ 
\multirow{2}{*}{$HM_{V}$} &  \multirow{2}{*}{$l_{HM}\times w_{HM} \times t_{HM}$} & \multirow{2}{*}{$100nm \times 50nm \times 3nm$}  \\ 
{}&{}&{} \\ 
{$RA$}&    {resistance-area product} & { 10 $\Omega.\mu m^2$}  \\
{$V_{0}$}&    {Fitting parameter} & {0.65}  \\ 
{$TMR_0$}&    {tunneling magnetoresistance} & {100}  \\ \hline
\end{tabular}
\end{table}

\begin{figure}[!t]
\centering
\includegraphics[width=3.4in]{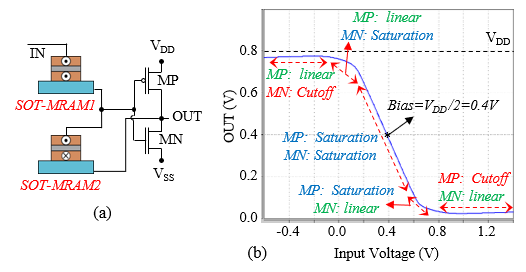}
\caption{(a) The SOT-MRAM based neuron, (b) The VTC curves showing various operating regions of PMOS (MP) and NMOS (MN) transistors.} 
\label{fig:neuron}
\end{figure} 

\subsection{SOT-MRAM Based Neuron}

Fig. \ref{fig:neuron} (a) shows the bitcell structure of the proposed neuron, which includes two SOT-MRAM devices and a CMOS-based inverter. The magnetization configurations of SOT-MRAM1 and SOT-MRAM2 devices should be in $P$ and $AP$ states, respectively. The SOT-MRAMs in the neuron's circuit create a voltage divider, which reduces the slope of the linear operating region in the inverter's voltage transfer characteristic (VTC) curve. The reduction in the slope of the linear region in the CMOS inverter creates a smooth high-to-low output voltage transition, which enables the realization of a $sigmoid$ activation function. Fig. \ref{fig:neuron} (b) shows the SPICE circuit simulation results of the proposed SOT-MRAM based neuron using $V_{DD}=0.8V$ and $V_{SS}=0V$. The results verify that the neuron can approximate a $sigmoid (-x)$ activation function that is biased around $b=\frac{1}{2}(V_{DD}-V_{SS})$ voltage. The non-zero bias voltage can be canceled at both circuit- and algorithm-level, as described in the next sections.




\begin{figure}[!t]
\centering
\includegraphics[scale=0.9]{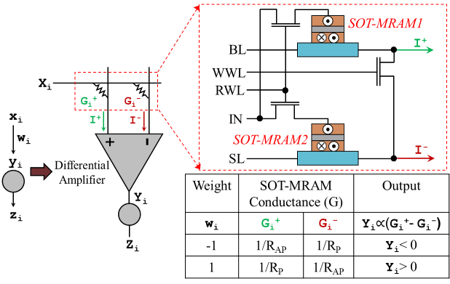}
\caption{The SOT-MRAM based binary synapse.} 
\label{fig:synapse}
\end{figure}

\subsection{SOT-MRAM Based Synapse}
SOT-MRAM cells are capable of realizing two resistive levels, i.e. $R_P$ and $R_{AP}$. The combination of two SOT-MRAM cells and a differential amplifier can produce the positive and negative weights required for the implementation of a binary synapse. Fig. \ref{fig:synapse} shows a neuron with $Y_i=X_i \times W_i$ as its input, in which $X_i$ is the input signal and $W_i$ is a binarized weight. The corresponding circuit implementation is also shown in the figure, which includes two SOT-MRAM cells and a differential amplifier as the synapse. The output of the differential amplifier ($Y_i$) is proportional to ($I^+-I^-$), where $I^+ = X_iG_i^+$ and $I^- = X_iG_i^-$. Thus, $Y_i \propto X_i(G_i^+-G_i^-$), in which $G_i^+$ and $G_i^-$ are the conductance of SOT-MRAM1 and SOT-MRAM2, respectively. The conductance of SOT-MRAMs can be adjusted to realize negative and positive weights in a binary synapse. For instance, for $\textbf{W}_i=-1$, SOT-MRAM1 and SOT-MRAM2 should be in $AP$ and $P$ states, respectively. According to Eq. (\ref{EqR}) $R_{AP}>R_P$, which means $G_{AP}<G_P$ since $G=1/R$, therefore $G_i^+ < G_i^-$ and $Y_i<0$.


\begin{figure*}[!t]
\centering
\includegraphics[width=7in]{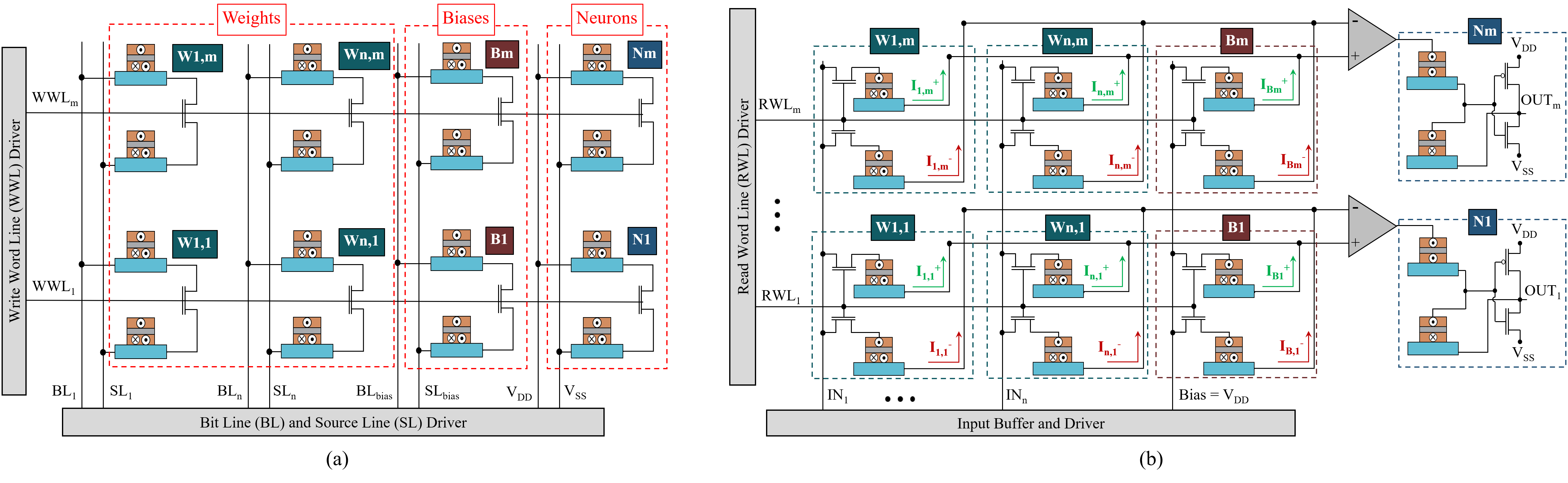}
\vspace{-3mm}
\caption{An $n \times m$ SOT-MRAM based single-layer perceptron. (a) The training path, and (b) inference path.}
\label{fig:arch}
\end{figure*}

\section{Proposed SOT-MRAM based MLP Architecture}
Fig. \ref{fig:arch} exhibit the training and inference paths of a proposed $n \times m$ SOT-MRAM based single layer perceptron, which are shown separately for simplicity. The synaptic connections are designed in the form of a crossbar architecture, in which the number of columns and rows are defined based on the number of nodes in input and output layers, respectively. During the training phase, the resistance of the SOT-MRAM based synapses will be tuned using the bit-lines (BLs) and source-lines (SLs) which are shared among different rows, as shown in Fig. \ref{fig:arch} (a). The write word line (WWL) control signals will only activate one row in each clock cycle, thus the entire array can be updated using $j$ clock cycles, where $j$ is equal to the number of neurons in the output layer. Moreover, to tune the states of the SOT-MRAMs in the neurons according to requirements mentioned Section II.A, the BL and SL control signals for the neuron are set to VDD and VSS, respectively, as shown in Fig. \ref{fig:arch} (a).


In the inference phase, the BL and SL control signals are in high-impedance (Hi-Z) state, and read word line (RWL) and WWL control signals are connected to VDD and GND, respectively. This will stop the write operation in synapses, and generate $I^+$ and $I^-$ currents shown in Fig. \ref{fig:arch} (b). The amplitude of produced currents depends on the input ($IN$) signals and the resistances of SOT-MRAM synapses. Each row includes a shared differential amplifier, which generates an output voltage proportional to $\sum_{i} (I_{i,n}^+-I_{i,n}^-)$ for the $n$th row, where $i$ is the total number of nodes in the input layer. Finally, the outputs of the differential amplifiers are connected to the SOT-MRAM based sigmoidal neurons. The entire inference operation occurs in parallel and in a single clock cycle. The required signaling to control the training and inference operations is listed in Table \ref{tab:signaling}. One of the main advantages of the proposed SOT-MRAM based perceptron architecture is that it can be readily concatenated to form an MLP classifier, which can still operate in a single clock cycle as it will be shown in Section V.

\begin{table}[!t]
\caption{The required signaling to control the proposed SOT-MRAM based perceptron array.}
\label{tab:signaling}
\centering
\begin{tabular}{lcccccc}
\hline
\multicolumn{2}{l}{\textbf{Operation}}    & \textbf{WWL} & \textbf{RWL} & \textbf{BL}   & \textbf{SL}   & \textbf{IN}   \\ \hline
\multirow{2}{*}{Training} & $\textbf{W}_i=+1$ & VDD & GND & VDD  & GND  & Hi-Z \\
                          & $\textbf{W}_i=-1$ & VDD & GND & GND  & VDD  & Hi-Z \\ \hline
\multicolumn{2}{l}{Inference}    & GND & VDD & Hi-Z & Hi-Z & VIN  \\ \hline
\end{tabular}
\end{table}

\section{Hardware-aware Learning Mechanism}

To train the proposed SOT-MRAM based MLP classifier, a hardware-aware learning mechanism should be developed which incorporates the characteristics and limitations of our SOT-MRAM based neurons and synapses. Herein, we use a two-stage teacher-student approach, in which both teacher and student networks have identical topologies. Table \ref{tab:learning} provides the notations and descriptions for teacher and student networks, in which $x$ is the input of the network and $y_i$ and $o_i$ are the input and output of the $i$th neuron, respectively.

To incorporate the features of the SOT-MRAM based synapses and neurons within our training mechanism, we have made two modifications to the approaches previously used for training binarized neural networks \cite{xnor-net,binaryconnect}. First, we have used binarized biases in the student networks instead of real-valued biases. Second, since our SOT-MRAM neuron realizes real-valued sigmoidal activation function ($sigmoid(-x)$) without any computation overheads, we could avoid binarizing the activation functions and reduce the possible information loss in the teacher or student networks \cite{xnor-net}. Herein, after each weight update in the teacher network we clip the real-valued weights within the $[-1,1]$ interval, and then use the below deterministic binarization approach to binarize the weights:

\begin{equation} 
\small
\label{Eq:deter_bin} 
W_{ij} = 
\begin{cases} 
+1, & \bar w_{ij} \ge \Delta_B  \\ 
-1, & \bar w_{ij}< \Delta_B
\end{cases} 
\end{equation}


\begin{table}[]
\caption{The notations and descriptions of the proposed learning mechanism for the SOT-MRAM based MLP.}
\label{tab:learning}
\centering
\begin{tabular}{lcc}
\hline
\multirow{2}{*}{}                                             & \multicolumn{1}{c}{\multirow{2}{*}{Teacher Network}} & \multicolumn{1}{c}{\multirow{2}{*}{Student Network}}                                                                      \\
                                                              & \multicolumn{1}{c}{}                                 & \multicolumn{1}{c}{}                                                                                                      \\ \hline
Weights                                                       & $\textbf{W}_i \in R$                                 & $\textbf{W}_i \in \{-1, +1\}$\\ \hline
Biases                                                        & $\textbf{B}_i \in R$                                 & $\textbf{W}_i \in \{-1, +1\}$ \\ \hline
\begin{tabular}[c]{@{}l@{}}Transfer Function\end{tabular}  & $y_i=\textbf{w}_i x + \textbf{b}_i$
& $y_i=\textbf{w}_i x + \textbf{b}_i$                                                                                                                \\ \hline
\begin{tabular}[c]{@{}l@{}}Activation Function\end{tabular} & $o_i=sigmoid(-y_i)$                                      & $o_i=sigmoid(-y_i)$                                                                                                           \\ \hline
\end{tabular}
\end{table}

\noindent where $\Delta_B=0$ is threshold parameters for binarized weights. Finally, once all the binarized weights are trained we will use a mapping mechanism to convert them to resistive states in SOT-MRAM based synapses according to Section II.B. It is worth noting that, the stochastic binarization \cite{binaryconnect} scheme can also be used to quantize the weights and biases. However, the stochastic rounding approach exhibits its advantages in deeper neural networks which are not the focus of this paper. In fact, we initially leveraged stochastic rounding in our simulations and while the training times were approximately 10-fold longer, the obtained accuracy values were comparable to those realized by the deterministic rounding approach. 

\section{Simulation Results}
\subsection{Circuit-Level Simulation of SOT-MRAM based Neuron}
Herein, we used the SPICE circuit simulator to measure the power consumption of our proposed SOT-MRAM based sigmoid neuron. The results obtained show the average power consumption of $64 \mu W$ for the SOT-MRAM based sigmoid neuron. Moreover, the layout design of the proposed neuron circuit shows an area consumption of $13\lambda \times 30\lambda$, in which $\lambda$ is a technology-dependent parameter. Herein, we used the 14nm FinFET technology, which leads to the approximate area consumption of $0.02 \mu m^2$. Table \ref{tab:comparison} provides a comparison between our SOT-MRAM based sigmoidal neuron and previous power- and area-efficient analog neurons \cite{neuron1,neuron2}.

To provide a fair comparison in terms of area and power dissipation, we have utilized the general scaling method \cite{Stillmaker2017Scaling7nm} to normalize the power dissipation and area of the designs listed in Table \ref{tab:comparison}. Voltage and area scale at different rates of $U=\frac{0.8}{VDD_x}$ and $S=\frac{14nm}{tech-node}$, respectively, where $VDD_x$ and $tech-node$ are the nominal voltage and technology node used in the studied neuron designs. It shall be noted that we used 0.8 (V) nominal voltage and 14nm FinFET technology in our design. Moreover, power and area consumption values are scaled with respect to $1/U^2$ and $1/S^2$, respectively \cite{Stillmaker2017Scaling7nm}.   The results obtained exhibit that the proposed SOT-MRAM based neuron achieves significant area reduction, while realizing comparable power consumption compared to the existing power- and area-efficient analog neuron implementations. This leads to a $74\times$ and $12\times$ reduction in power-area product compared to the designs introduced in \cite{neuron1} and \cite{neuron2}, respectively.  
   

\begin{table}[]
\centering
\caption{Performance comparison for various analog sigmoidal neuron implementations.}
\label{tab:comparison}
\begin{tabular}{lccc}
\hline
                   & \cite{neuron1} & \cite{neuron2} & Proposed Herein \\ \hline
Power Consumption  & 7.4$\times$      & 0.98$\times$     & 1$\times$                                                        \\
Area Consumption   & 10$\times$       & 12.3$\times$     & 1$\times$                                                        \\ \hline
Power-Area Product & 74$\times$       & 12$\times$       & 1$\times$                                                        \\ \hline
\end{tabular}
\end{table}

\subsection{Architecture-level Simulation}
\begin{figure}[!t]
\centering
\includegraphics[width=3.4in]{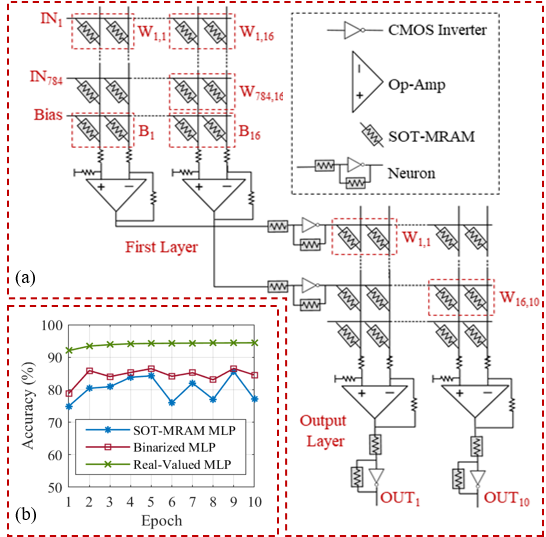}
\caption{(a) The $784 \times 16 \times 10$ SOT-MRAM based MLP circuit, (b) Accuracy comparison for MNIST application using a $784\times16\times10$ MLP. }
\label{fig:mlp}
\end{figure}

Herein, we developed a Python-based simulation framework based on \cite{zandjetc2019} to realize the SPICE implementation of our SOT-MRAM based MLP classifier. Fig. \ref{fig:mlp} (a) depicts the circuit realization of a $784\times16\times10$ SOT-MRAM based MLP classifier. A comparison between the MNIST \cite{MNIST} classification accuracy of SOT-MRAM MLP classifier and conventional real-valued and binarized MLP architectures is shown in Fig. \ref{fig:mlp} (b). The results show a comparable maximum classification accuracy of 86.54\% and 85.56\% in the first 10 epochs for binarized and SOT-MRAM based MLP classifiers, respectively. 

Moreover, Table \ref{tab:archcompare} provides a comparison between the analog IMC-based MLP classifier proposed herein and various hardware implementations of a $784\times16\times10$ binarized MLP architecture. As listed in the table, our analog MLP classifier completes the recognition task in a single clock cycle, while a highly-parallel digital implementation on GPU and a high-performance mixed-signal IMC architecture require at least $10^5$ and $10^2$ clock cycles, respectively, to complete the similar task. It is worth noting that digital CPU or GPU implementations can support higher clock frequencies as listed in the table. However, the difference between total clock cycles is so large that our analog IMC realization can still achieve at least four and five orders of magnitude performance improvement compared to GPU and CPU implementation, respectively.




\begin{table}[]
\caption{Performance comparison among various implementations of the binarized $784 \times 16 \times 10$ MLP classifier.}
\centering
\begin{threeparttable}
\begin{tabular}{cclccc}
\hline
\multirow{2}{*}{Architecture} & \multicolumn{3}{c}{Domain}                  & \multirow{2}{*}{\begin{tabular}[c]{@{}c@{}}Frequency\\ (GHz)\end{tabular}} & \multirow{2}{*}{Total Clocks}               \\ \cline{2-4}
                              & \multicolumn{2}{c}{MAC}     & Act. Func.    &                                                                            &                                             \\ \hline
CPU                           & \multicolumn{2}{c}{Digital} & Digital       & 3.7\tnote{(1)}                                                                        & $10^7-10^8$ \tnote{(*)}  \\
GPU                           & \multicolumn{2}{c}{Digital} & Digital       & 1.35\tnote{(2)}                                                                       & $10^5-10^6$ \tnote{(*)} \\
IMC \cite{inmemory-fan}                          & \multicolumn{2}{c}{Digital} & Digital       & 0.667\tnote{(3)}                                                                    & $10^4-10^5$ \tnote{(*)} \\
IMC \cite{in-memory-PCM}                          & \multicolumn{2}{c}{Analog}  & Digital       & 0.2-0.667\tnote{(4)}                                                                  & $10^2-10^4$ \tnote{(*)} \\
\textbf{Proposed Here }                     & \multicolumn{2}{c}{\textbf{Analog}}  & \textbf{Analog}        & \textbf{0.2-0.667}\tnote{(4)}                                                                  & \textbf{1}                                           \\ \hline
\end{tabular}
\begin{tablenotes}
\small
\item[(1)] \footnotesize Implemented on Intel Core i9-10900X.
\item[(2)] \footnotesize Implemented on NVIDIA GeForce RTX 2080 Ti.
\item[(3)] \footnotesize Not reported in \cite{inmemory-fan}. Estimated according to Everspin STT-MRAM.    
\item[(4)] \footnotesize Not reported in \cite{in-memory-PCM}. Clock frequency is expected to be reduced due to the parasitic effects in analog domain.
\item[(*)] \footnotesize We have reported a range instead of exact values to compensate for possible variations in users' programming and implementation skills. 
\end{tablenotes} 
\end{threeparttable}
\label{tab:archcompare}
\end{table}

\section{Conclusion}
In this paper, we proposed a power- and area-efficient SOT-MRAM based sigmoidal neuron, which was leveraged along-with SOT-MRAM based binary synapses to construct an analog IMC architecture for MLP classifiers. The developed neuron and synapse bitcells could be implemented within the same memory subarray, enabling a single-cycle operation for the analog IMC-based MLP architecture while removing the need for signal conversion units. We implemented a $784\times16\times10$ SOT-MRAM based MLP using the SPICE circuit simulator and compared its performance with various hardware realizations of an MLP classifier. The results exhibited at least two orders of magnitude increase in the processing speed of our analog IMC architecture compared to the highest performance MLP classifier implemented on a mixed-signal analog/digital IMC architecture.

\bibliographystyle{IEEEtran}

\balance
\bibliography{ref}
%



\end{document}